\documentclass[twocolumn,showpacs,amsmath,amssymb]{revtex4}

\begin{document}
\title{Analytical formula connecting entangled state
and the closest disentangled state}
\author{Satoshi Ishizaka}
%\email{isizaka@frl.cl.nec.co.jp}
\affiliation{Fundamental Research Laboratories, NEC Corporation,
34 Miyukigaoka, Tsukuba, Ibaraki, 305-8501 Japan}
\date{\today}
%%%%%%%%%%%%%%%%%%%%%%%%%%%%%%%%%%%%%%%%%%%%%%%%%%%%%%%%%%%%%%%%%%%%%%%%%%%%%%%
\begin{abstract}
The separable state closest to a given entangled state
in the relative entropy measure is called the closest disentangled state.
We provide an analytical formula connecting the entangled state
and the closest disentangled state in two qubits.
Using this formula, when any disentangled state ($\sigma$) located at the
entangle-disentangle boundary is given, entangled states to which
$\sigma$ is closest can be obtained analytically.
Further, this formula naturally defines the direction normal to
the boundary surface.
The direction is uniquely determined by $\sigma$ in almost all cases.
\end{abstract}
\pacs{03.67.-a, 03.65.Ud}
%%%%%%%%%%%%%%%%%%%%%%%%%%%%%%%%%%%%%%%%%%%%%%%%%%%%%%%%%%%%%%%%%%%%%%%%%%%%%%%
\maketitle
\setlength{\baselineskip}{3.85mm}
%%%%%%%%%%%%%%%%%%%%%%%%%%%%%%%%%%%%%%%%%%%%%%%%%%%%%%%%%%%%%%%%%%%%%%%%%%%%%%%
%%
%%%%%%%%%%%%%%%%%%%%%%%%%%%%%%%%%%%%%%%%%%%%%%%%%%%%%%%%%%%%%%%%%%%%%%%%%%%%%%%
Quantum entanglement is the most striking feature of quantum mechanics.
Several measures have been proposed to quantify
the amount of entanglement \cite{Horodecki01a}.
One of the important measures is
relative entropy of entanglement \cite{Vedral97a,Vedral98a},
which is defined as
%%%%%%%%%%%%%%%%%%%%%%%%%%%%%%%%%%%%%%%%%%%%%%%%%%%%%%%%%%%%%%%%%%%%%%%%%%%%%%%
\begin{equation}
E_R(\varrho)=\min_{\sigma \in {\cal D}} S(\varrho||\sigma)
=\min_{\sigma \in {\cal D}}\big[\hbox{Tr}\varrho\ln \varrho-
\hbox{Tr}\varrho\ln \sigma\big],
\end{equation}
%%%%%%%%%%%%%%%%%%%%%%%%%%%%%%%%%%%%%%%%%%%%%%%%%%%%%%%%%%%%%%%%%%%%%%%%%%%%%%%
where the minimization is performed over all density matrices
in the set of disentangled states
($\ln$ denotes the natural logarithm throughout this paper).
The states achieving the minimum are called the closest disentangled states.
The efficient algorithm for the numerical minimization has been
proposed in \cite{Rehacek02a}.
However, the analytical minimization is difficult in general,
and the analytical solutions only for several classes of states have been shown
so far.
\par
%%%%%%%%%%%%%%%%%%%%%%%%%%%%%%%%%%%%%%%%%%%%%%%%%%%%%%%%%%%%%%%%%%%%%%%%%%%%%%%
In this paper, 
we provide an analytical formula connecting entangled states
and the closest disentangled state in two qubits.
Using this formula, when any disentangled state ($\sigma$) located at the
entangle-disentangle boundary is given, entangled states $\varrho$ to which
$\sigma$ is closest can be obtained analytically.
Further, this formula naturally defines the direction normal to
the boundary surface when relative entropy is regarded as the distant measure.
The direction is uniquely determined by $\sigma$ in almost all
cases.
The discussions in this paper are limited to the case of two qubits.
\par
%%%%%%%%%%%%%%%%%%%%%%%%%%%%%%%%%%%%%%%%%%%%%%%%%%%%%%%%%%%%%%%%%%%%%%%%%%%%%%%
The strategy of the minimization is based on the following fact:
{\it 
Any full rank density matrix can be transformed into
a unique Bell diagonal state by applying
a suitable local filtering operation}
\cite{Verstraete01a,Cen02a}.
The explicit formulation using the Lorentz matrices has been shown in 
\cite{Verstraete01a}, and that using the Wootters basis
\cite{Wootters98a}
has been shown in \cite{Cen02a}.
The entanglement manipulation by local filtering on a single copy
was originally considered in \cite{Linden98a,Kent99a}.
Further, the connection between the Wootters basis and Lewenstein-Sanpera
decomposition \cite{Lewenstein98a} has been shown in 
\cite{Akhtarshenas02a}.
According to these results, any full rank density matrix can be written as
%%%%%%%%%%%%%%%%%%%%%%%%%%%%%%%%%%%%%%%%%%%%%%%%%%%%%%%%%%%%%%%%%%%%%%%%%%%%%%%
$
\varrho\!=\!\frac{1}{N}(F_A \otimes F_B)\varrho_{\rm BD}
(F_A^\dagger \otimes F_B^\dagger)
$,
%%%%%%%%%%%%%%%%%%%%%%%%%%%%%%%%%%%%%%%%%%%%%%%%%%%%%%%%%%%%%%%%%%%%%%%%%%%%%%%
where
$N\!=\!\hbox{Tr}[(F_A \otimes F_B )\varrho_{\rm BD}(F_A^\dagger \otimes F_B^\dagger )]$
and $\varrho_{\rm BD}\!=\!\sum_{i=0}^3 p_i|e_i\rangle\langle e_i|$
is the Bell diagonal state.
%%%%%%%%%%%%%%%%%%%%%%%%%%%%%%%%%%%%%%%%%%%%%%%%%%%%%%%%%%%%%%%%%%%%%%%%%%%%%%%
$\varrho_{\rm BD}$ can be always chosen in a canonical form, in which
$p_0$ is maximum ($p_0\!\ge\!p_1,p_2,p_3$) and $|e_i\rangle$ is a
fixed set of Bell basis, since suitable local unitary operations transform
any Bell diagonal state into the canonical form \cite{Note1,Horodecki96b}.
On the other hand, $\varrho$ can be always expressed using the Wootters
basis as
$\varrho\!=\!\sum_{i=0}^{3}\lambda_i|\phi_i\rangle\langle\phi_i|$
\cite{Wootters98a},
where $\langle \phi_i|\tilde\phi_j\rangle
\!=\!\langle \tilde \phi_i|\phi_j\rangle\!=\!\delta_{ij}$ defining
the tilde operation as
$|\tilde\psi\rangle\!=\!(\sigma_2\otimes\sigma_2)|\psi^*\rangle$.
The Bell basis is one of the Wootters basis by itself 
(adding a suitable global phase if necessary),
and hence $|\tilde e_i\rangle\!=\!|e_i\rangle$.
The concurrence of $\varrho$ is
$C\!=\!\lambda_0-\lambda_1-\lambda_2-\lambda_3$.
Although the actual definition of the concurrence is
$\max(\lambda_0-\lambda_1-\lambda_2-\lambda_3,0)$
\cite{Wootters98a},
we define it here such that $C$ can take negative when $\varrho$ is
disentangled.
Further, we can put $\det F_A\!=\!\det F_B\!=\!1$ without loss of generality.
In this choice, $|\phi_i\rangle\!=\!(F_A\!\otimes\!F_B)|e_i\rangle$
(using $A^T\sigma_2 A\!=\!(\det A)\sigma_2$,
it can be checked that
$|\phi_i\rangle\!=\!(F_A\!\otimes\!F_B)|e_i\rangle$
satisfy $\langle \phi_i|\tilde\phi_j\rangle\!=\!\delta_{ij}$)
and $\lambda_i\!=\!p_i/N$
\cite{Verstraete01a,Cen02a}.
\par
%%%%%%%%%%%%%%%%%%%%%%%%%%%%%%%%%%%%%%%%%%%%%%%%%%%%%%%%%%%%%%%%%%%%%%%%%%%%%%%
Considering these facts carefully, the following structure
of the Hilbert space is revealed:
The total 15-dimensional space is divided into subspaces
in such a way that
only the states in the same subspace can be transformed to each other by local
filtering
(we only consider the full rank density matrices for the sake of simplicity
\cite{Verstraete01a,Cen02a}).
Each subspace contains the canonical Bell diagonal states
having the same $p_i$ only (we do not care about the exchange
among $p_1$, $p_2$, and $p_3$).
Therefore, the subspaces can be specified by three parameters
($p_i$ with $\sum p_i\!=\!1$), and each subspace constitutes a 12-dimensional
manifold.
Further, since $C\!>\!0$ for entangled states
and $C\!\le\!0$ for disentangled states,
and the local filtering does not change
the sign of $C$ (see Theorem 1 in \cite{Verstraete01a}),
we can define {\it boundary subspaces}, in which all the states have just
zero of $C$.
The boundary subspaces are specified by two parameters,
and the assemble of the boundary subspaces constitutes the 14-dimensional
boundary surface separating the region of entangled states and disentangled
states.
\par
%%%%%%%%%%%%%%%%%%%%%%%%%%%%%%%%%%%%%%%%%%%%%%%%%%%%%%%%%%%%%%%%%%%%%%%%%%%%%%%
Then, we perform the minimization of $S(\varrho||\sigma)$ separately:
First we obtain the extremal conditions by minimizing within a
fixed subspace, and second we obtain the conditions to determine
the boundary subspace containing the closest disentangled state.
Finally we solve the equations obtained.
Hereafter, the entangled state and the corresponding closest disentangled state
is denoted by $\varrho$ and $\sigma$, respectively.
Further, we first assume that $\sigma$ is full rank.
The case of $\sigma$ with lower rank will be discussed in the last part of
this paper.
\par
%%%%%%%%%%%%%%%%%%%%%%%%%%%%%%%%%%%%%%%%%%%%%%%%%%%%%%%%%%%%%%%%%%%%%%%%%%%%%%%
We have already performed the first step and obtained the set of equations
$\sigma$ must satisfy \cite{Ishizaka02a}.
We briefly repeat the results here.
The states obtained from $\sigma$ by local filtering belong to the same
subspace as $\sigma$ by its definition.
Let us consider Bob's local filtering of
$\sigma'\!=\!\frac{1}{N}
(I \otimes e^{t\vec n \cdot \vec \sigma/2})
\sigma
(I \otimes e^{t\vec n \cdot \vec \sigma/2})$,
where $t$ is a real parameter and $\vec\sigma\!=\!(\sigma_1,\sigma_2,\sigma_3)$
is the vector of Pauli matrices.
In order that $S(\varrho||\sigma)$ is minimum, the linear coefficient
of $t$ in the expansion of $\hbox{Tr}\varrho \ln \sigma'$ must be zero.
As the result, using
$\ln(X+tY)\!=\!\ln X
+t\int_0^\infty\!\!\!\frac{1}{X+z}Y\frac{1}{X+z}dz+\cdots$
we obtained
%%%%%%%%%%%%%%%%%%%%%%%%%%%%%%%%%%%%%%%%%%%%%%%%%%%%%%%%%%%%%%%%%%%%%%%%%%%%%%%
\begin{equation}
\hbox{Tr} (I\otimes \sigma_i)Z\sigma+
\hbox{Tr} Z(I\otimes \sigma_i)\sigma=0
\hbox{~~~for $i=1,2,3$}
\label{eq: equations by local filter}
\end{equation}
%%%%%%%%%%%%%%%%%%%%%%%%%%%%%%%%%%%%%%%%%%%%%%%%%%%%%%%%%%%%%%%%%%%%%%%%%%%%%%
where
%%%%%%%%%%%%%%%%%%%%%%%%%%%%%%%%%%%%%%%%%%%%%%%%%%%%%%%%%%%%%%%%%%%%%%%%%%%%%%%
\begin{equation}
Z\equiv
\int_0^\infty \!\!\! \frac{1}{\sigma+z}\varrho\frac{1}{\sigma+z}dz-I
\label{eq: Z}
\end{equation}
%%%%%%%%%%%%%%%%%%%%%%%%%%%%%%%%%%%%%%%%%%%%%%%%%%%%%%%%%%%%%%%%%%%%%%%%%%%%%%%
(Eq.\ (\ref{eq: equations by local filter}) holds for $i\!=\!0$ also, but
we omitted it since
$\hbox{Tr}Z\sigma\!=\!0$ is obvious from Eq.\ (\ref{eq: Z})).
In the same manner, considering Bob's local unitary transformation of
$\sigma'\!=\!
(I \otimes e^{i t\vec n \cdot \vec \sigma/2})
\sigma
(I \otimes e^{-i t\vec n \cdot \vec \sigma/2})$,
%%%%%%%%%%%%%%%%%%%%%%%%%%%%%%%%%%%%%%%%%%%%%%%%%%%%%%%%%%%%%%%%%%%%%%%%%%%%%%%
we obtained
%%%%%%%%%%%%%%%%%%%%%%%%%%%%%%%%%%%%%%%%%%%%%%%%%%%%%%%%%%%%%%%%%%%%%%%%%%%%%%%
\begin{equation}
\hbox{Tr} (I\otimes \sigma_i)Z\sigma-
\hbox{Tr} Z(I\otimes \sigma_i)\sigma=0
\hbox{~~~for $i=1,2,3$}.
\label{eq: equations by local unitary}
\end{equation}
%%%%%%%%%%%%%%%%%%%%%%%%%%%%%%%%%%%%%%%%%%%%%%%%%%%%%%%%%%%%%%%%%%%%%%%%%%%%%%%
From Eqs.\ (\ref{eq: equations by local filter}) and
(\ref{eq: equations by local unitary}), we have
\begin{equation}
\hbox{Tr} (I\otimes \sigma_i)Z\sigma=0,
\label{eq: equations by Bob's local filter}
\end{equation}
%%%%%%%%%%%%%%%%%%%%%%%%%%%%%%%%%%%%%%%%%%%%%%%%%%%%%%%%%%%%%%%%%%%%%%%%%%%%%%%
which is sufficient for
$\hbox{Tr}Z(I\otimes \sigma_i) \sigma \!=\!0$
since $Z$ is Hermitian.
%%%%%%%%%%%%%%%%%%%%%%%%%%%%%%%%%%%%%%%%%%%%%%%%%%%%%%%%%%%%%%%%%%%%%%%%%%%%%%%
Therefore, $\sigma$ must satisfy
Eqs.\ (\ref{eq: equations by Bob's local filter}) and Alice's
counterparts
%%%%%%%%%%%%%%%%%%%%%%%%%%%%%%%%%%%%%%%%%%%%%%%%%%%%%%%%%%%%%%%%%%%%%%%%%%%%%%%
\begin{equation}
\hbox{Tr} (\sigma_i\otimes I)Z\sigma=0.
\label{eq: equations by Alice's local filter}
\end{equation}
%%%%%%%%%%%%%%%%%%%%%%%%%%%%%%%%%%%%%%%%%%%%%%%%%%%%%%%%%%%%%%%%%%%%%%%%%%%%%%%
These are fundamental equations for the closest
disentangled state
(and even for the closest positive partial transposed states),
because these equations must hold in any multi-party
systems and any higher dimensional systems \cite{Ishizaka02a}.
\par
%%%%%%%%%%%%%%%%%%%%%%%%%%%%%%%%%%%%%%%%%%%%%%%%%%%%%%%%%%%%%%%%%%%%%%%%%%%%%%%
The second step is to obtain the conditions to
determine the boundary subspace containing $\sigma$.
The canonical Bell diagonal state in the boundary subspace must have zero of
$C\!=\!p_0-p_1-p_2-p_3$,
but $\sum p_i\!=\!1$ due to the normalization.
As the result, $p_0$ must be $1/2$
(this is also obvious from the fact that the concurrence of the
Bell diagonal state is $2p_0-1$ \cite{Bennett96a,Wootters98a}).
Therefore, $\sigma$, which belongs to the boundary subspace,
must be obtained by local filtering from 
$\sigma_{\rm BD}\!=\!
\frac{1}{2}|e_0\rangle\langle e_0|+\sum_{i=1}^3 p_i|e_i\rangle\langle e_i|$,
and hence $\sigma$ must be written as
%%%%%%%%%%%%%%%%%%%%%%%%%%%%%%%%%%%%%%%%%%%%%%%%%%%%%%%%%%%%%%%%%%%%%%%%%%%%%%%
\begin{eqnarray}
\sigma&=&\frac{1}{N}
(F_A \otimes F_B)
\sigma_{\rm BD}
(F_A^\dagger \otimes F_B^\dagger) \cr
&=&\frac{1}{2N}|\phi_0\rangle\langle \phi_0|+
\sum_{i=1}^3 \frac{p_i}{N}|\phi_i\rangle\langle \phi_i|,
\label{eq: parameterization}
\end{eqnarray}
%%%%%%%%%%%%%%%%%%%%%%%%%%%%%%%%%%%%%%%%%%%%%%%%%%%%%%%%%%%%%%%%%%%%%%%%%%%%%%%
where $N$ is again normalization, and the set of $|\phi_i\rangle$'s
is the Wootters basis of $\sigma$.
\par
%%%%%%%%%%%%%%%%%%%%%%%%%%%%%%%%%%%%%%%%%%%%%%%%%%%%%%%%%%%%%%%%%%%%%%%%%%%%%%%
Adopting this parameterization, to determine the subspace containing $\sigma$
is equivalent to determine $p_1$, $p_2$, and $p_3$ under the constraint
of $p_1+p_2+p_3\!=\!1/2$ and $0\!\le\!p_i\!\le\!1/2$.
However, we assumed that $\sigma$ is full rank,
which ensures that the minimum of $S(\varrho||\sigma)$ is
achieved at $0\!<\!p_i\!<\!1/2$.
Therefore, we can simply minimize 
%%%%%%%%%%%%%%%%%%%%%%%%%%%%%%%%%%%%%%%%%%%%%%%%%%%%%%%%%%%%%%%%%%%%%%%%%%%%%%%
\begin{equation}
f\equiv-\hbox{Tr}\rho\ln\sigma+l \big(p_1+p_2+p_3-1/2 \big),
\end{equation}
%%%%%%%%%%%%%%%%%%%%%%%%%%%%%%%%%%%%%%%%%%%%%%%%%%%%%%%%%%%%%%%%%%%%%%%%%%%%%%%
with $l$ being a Lagrange multiplier.
Considering the small change of $p_i$ (note that $N$ also contains $p_i$),
we obtain
%%%%%%%%%%%%%%%%%%%%%%%%%%%%%%%%%%%%%%%%%%%%%%%%%%%%%%%%%%%%%%%%%%%%%%%%%%%%%%%
\begin{equation}
\frac{1}{N}\hbox{Tr}
Z|\phi_i\rangle\langle \phi_i|=l \hbox{~~~for $i=1,2,3$}.
\end{equation}
%%%%%%%%%%%%%%%%%%%%%%%%%%%%%%%%%%%%%%%%%%%%%%%%%%%%%%%%%%%%%%%%%%%%%%%%%%%%%%%
Further, multiplying $p_i$ on both side of the above equations
and adding them, we find
%%%%%%%%%%%%%%%%%%%%%%%%%%%%%%%%%%%%%%%%%%%%%%%%%%%%%%%%%%%%%%%%%%%%%%%%%%%%%%%
\begin{equation}
\sum_{i=1}^3\frac{p_i}{N}\hbox{Tr}
Z|\phi_i\rangle\langle \phi_i|=
\hbox{Tr}Z\sigma-\frac{1}{2N}\hbox{Tr}Z|\phi_0\rangle\langle\phi_0|=
\frac{l}{2},
\end{equation}
%%%%%%%%%%%%%%%%%%%%%%%%%%%%%%%%%%%%%%%%%%%%%%%%%%%%%%%%%%%%%%%%%%%%%%%%%%%%%%%
and we finally obtain the extremal conditions as
%%%%%%%%%%%%%%%%%%%%%%%%%%%%%%%%%%%%%%%%%%%%%%%%%%%%%%%%%%%%%%%%%%%%%%%%%%%%%%%
\begin{equation}
\langle\phi_i|Z|\phi_i\rangle+\langle\phi_0|Z|\phi_0\rangle=0
\hbox{~~~for $i=1,2,3$}.
\label{eq: equations by subspace}
\end{equation}
%%%%%%%%%%%%%%%%%%%%%%%%%%%%%%%%%%%%%%%%%%%%%%%%%%%%%%%%%%%%%%%%%%%%%%%%%%%%%%%
There are three equations in the above, in spite that the boundary
subspaces are specified by two parameters as mentioned before.
This is because Eqs.\ (\ref{eq: equations by subspace}) includes the
trace condition of $\hbox{Tr}\varrho\!=\!\hbox{Tr}\sigma$.
\par
%%%%%%%%%%%%%%%%%%%%%%%%%%%%%%%%%%%%%%%%%%%%%%%%%%%%%%%%%%%%%%%%%%%%%%%%%%%%%%%
Then, all equations of the extremal conditions are
Eqs. (\ref{eq: equations by Bob's local filter}),
(\ref{eq: equations by Alice's local filter}), and
(\ref{eq: equations by subspace}) with
$p_1+p_2+p_3\!=\!1/2$ (and thus
$\lambda_0-\lambda_1-\lambda_2-\lambda_3\!=\!0$).
The total number of these equations is 16 including the normalization
condition.
These must be solved for any given $\varrho$
in order to obtain a true closed formula for the closest disentangled state
(and for $E_R(\varrho)$).
However, since it is a hard problem unfortunately, we follow the
strategy in \cite{Rains99a}.
Namely, for any given $\sigma$, we obtain the set of $\varrho$ for
which $S(\varrho||\sigma)$ is minimum.
We can obtain it in an analytical form as shown below. 
\par
%%%%%%%%%%%%%%%%%%%%%%%%%%%%%%%%%%%%%%%%%%%%%%%%%%%%%%%%%%%%%%%%%%%%%%%%%%%%%%%
Let us assume that $\sigma$ with $C\!=\!0$ is given.
Here, we still assume that $\sigma$ is full rank.
It is convenient to express $\sigma$ in two representations:
%%%%%%%%%%%%%%%%%%%%%%%%%%%%%%%%%%%%%%%%%%%%%%%%%%%%%%%%%%%%%%%%%%%%%%%%%%%%%%%
\begin{equation}
\sigma=\sum_{ij}\Lambda_{ij}|\phi_i\rangle\langle \phi_j|
=\sum_{ij}\Gamma_{ij}|i\rangle\langle j|,
\end{equation}
%%%%%%%%%%%%%%%%%%%%%%%%%%%%%%%%%%%%%%%%%%%%%%%%%%%%%%%%%%%%%%%%%%%%%%%%%%%%%%%
where $|\phi_i\rangle$'s and $|i\rangle$'s are the Wootters basis and
eigenstates of $\sigma$, respectively, and hence
$\Lambda_{ij}\!=\!\lambda_i\delta_{ij}$ and 
$\Gamma_{ij}\!=\!\gamma_i\delta_{ij}$ with 
$\gamma_i$'s being eigenvalues of $\sigma$.
These two basis sets are connected through a unitary matrix as
$\sqrt{\lambda_i}|\phi_i\rangle\!=\!\sum_j V_{ij}^*\sqrt{\gamma_j}|j\rangle$
\cite{Wootters98a}.
Since $|\phi_i\rangle$'s are non-orthogonal in general, it is also convenient
to introduce the matrices describing the non-orthogonality as follows:
%%%%%%%%%%%%%%%%%%%%%%%%%%%%%%%%%%%%%%%%%%%%%%%%%%%%%%%%%%%%%%%%%%%%%%%%%%%%%%%
\begin{equation}
Q_{ij}=\langle \phi_i|\phi_j\rangle \hbox{,~~~}
\Pi_{ij}=\langle \tilde \phi_i|\tilde \phi_j\rangle.
\label{eq: matrix Q}
\end{equation}
%%%%%%%%%%%%%%%%%%%%%%%%%%%%%%%%%%%%%%%%%%%%%%%%%%%%%%%%%%%%%%%%%%%%%%%%%%%%%%%
Using
the completeness of $I\!=\!\sum_i |\phi_i\rangle\langle \tilde \phi_i|
\!=\!\sum_i |\tilde \phi_i\rangle\langle \phi_i|$
\cite{Akhtarshenas02a,Cen02b} and
$I\!=\!\sum_{ij}Q_{ij}|\tilde \phi_i\rangle\langle \tilde \phi_j|
\!=\!\sum_{ij}\Pi_{ij}|\phi_i\rangle\langle \phi_j|$,
it is easy to check that 
Hermitian $Q$ and $\Pi$ satisfy
$QQ^T\!=\!I$, $\Pi=Q^*$, and $Q\Pi\!=\!\Pi Q\!=\!I$.
Further,
%%%%%%%%%%%%%%%%%%%%%%%%%%%%%%%%%%%%%%%%%%%%%%%%%%%%%%%%%%%%%%%%%%%%%%%%%%%%%%%
\begin{equation}
\Gamma=V^\dagger \sqrt{\Lambda}Q\sqrt{\Lambda}V
=U\sqrt{Q}\Lambda\sqrt{Q}U^\dagger,
\label{eq: matrix U}
\end{equation}
%%%%%%%%%%%%%%%%%%%%%%%%%%%%%%%%%%%%%%%%%%%%%%%%%%%%%%%%%%%%%%%%%%%%%%%%%%%%%%%
where we introduced the unitary $U$ by the singular value
decomposition of $U\sqrt{Q}\sqrt{\Lambda}V\!=\!\sqrt{\Gamma}$.
Then, we shall obtain $\varrho$ in the same representation as $\sigma$.
Namely, using $|\phi_i\rangle$ and $|i\rangle$ of $\sigma$,
%%%%%%%%%%%%%%%%%%%%%%%%%%%%%%%%%%%%%%%%%%%%%%%%%%%%%%%%%%%%%%%%%%%%%%%%%%%%%%%
\begin{equation}
\varrho=\sum_{ij}R^\Lambda_{ij}|\phi_i\rangle\langle \phi_j|
=\sum_{ij}R^E_{ij}|i\rangle\langle j|.
\end{equation}
%%%%%%%%%%%%%%%%%%%%%%%%%%%%%%%%%%%%%%%%%%%%%%%%%%%%%%%%%%%%%%%%%%%%%%%%%%%%%%%
Hereafter, the coefficient matrix in the Wootters basis and eigenstates
is denoted by suffix $\Lambda$ and $E$, respectively.
Two representations are connected through
$R^E\!=\!
\sqrt{\Gamma}V^\dagger\sqrt{\Lambda}R^\Lambda\sqrt{\Lambda}V\sqrt{\Gamma}
\!=\!U\sqrt{Q}R^\Lambda\sqrt{Q}U^\dagger$.
\par
%%%%%%%%%%%%%%%%%%%%%%%%%%%%%%%%%%%%%%%%%%%%%%%%%%%%%%%%%%%%%%%%%%%%%%%%%%%%%%%
Viewing the form of Eqs. (\ref{eq: equations by Bob's local filter}),
(\ref{eq: equations by Alice's local filter}), and
(\ref{eq: equations by subspace}),
it is found that,
when $\varrho$ is a solution,
$\varrho'\!=\!x\varrho+(1-x)\sigma$ is also a solution,
since
$Z\!\rightarrow\!xZ$ for $\varrho\!\rightarrow\!x\varrho+(1-x)\sigma$.
This implies that the normalization of $Z$ can be taken as a free parameter,
and we express $Z$ as
%%%%%%%%%%%%%%%%%%%%%%%%%%%%%%%%%%%%%%%%%%%%%%%%%%%%%%%%%%%%%%%%%%%%%%%%%%%%%%%
\begin{equation}
\langle i|Z| j \rangle
=\int_0^\infty\!\!\!\!\!
\frac{1}{\gamma_i+z}
R^E_{ij}
\frac{1}{\gamma_j+z}dz
-\delta_{ij}
\equiv x W^E_{ij}.
\end{equation}
%%%%%%%%%%%%%%%%%%%%%%%%%%%%%%%%%%%%%%%%%%%%%%%%%%%%%%%%%%%%%%%%%%%%%%%%%%%%%%%
Then, we have
$x W^E\!=\![R^E \circ \hat G ]-I$, where
$A\!\circ\!B$ is the Hadamard product defined as
$[A\circ B]_{ij}\!=\!A_{ij}B_{ij}$, 
$\hat G_{ij}\!=\!1/G_{ij}$, and
%%%%%%%%%%%%%%%%%%%%%%%%%%%%%%%%%%%%%%%%%%%%%%%%%%%%%%%%%%%%%%%%%%%%%%%%%%%%%%%
\begin{equation}
G_{ij}\equiv\left\{
\begin{array}{cl}
\gamma_i & \hbox{for $\gamma_i\!=\!\gamma_j$} \cr
\frac{\gamma_i-\gamma_j}{\ln \gamma_i-\ln \gamma_j}
         & \hbox{for $\gamma_i\!\ne\!\gamma_j$}.
\end{array}
\right.
\label{eq: matrix G}
\end{equation}
%%%%%%%%%%%%%%%%%%%%%%%%%%%%%%%%%%%%%%%%%%%%%%%%%%%%%%%%%%%%%%%%%%%%%%%%%%%%%%%
By inverting the above
(we can safely invert since $G_{ij}\!>\!0$ by $\gamma_i\!>\!0$),
we have
%%%%%%%%%%%%%%%%%%%%%%%%%%%%%%%%%%%%%%%%%%%%%%%%%%%%%%%%%%%%%%%%%%%%%%%%%%%%%%%
\begin{equation}
R^E=\Gamma+x[W^E \circ G],
\end{equation}
%%%%%%%%%%%%%%%%%%%%%%%%%%%%%%%%%%%%%%%%%%%%%%%%%%%%%%%%%%%%%%%%%%%%%%%%%%%%%%%
and therefore the problem is reduced to find $W^E$ satisfying the set of
all extremal conditions.
\par
%%%%%%%%%%%%%%%%%%%%%%%%%%%%%%%%%%%%%%%%%%%%%%%%%%%%%%%%%%%%%%%%%%%%%%%%%%%%%%%
For this purpose, it is convenient to express $Z$ in the $|\tilde\phi_i\rangle$
basis as $Z\!=\!\sum xW_{ij}|\tilde\phi_i\rangle\langle\tilde\phi_j|$.
Then, the extremal conditions of Eqs.\ (\ref{eq: equations by subspace})
become quite simple:
%%%%%%%%%%%%%%%%%%%%%%%%%%%%%%%%%%%%%%%%%%%%%%%%%%%%%%%%%%%%%%%%%%%%%%%%%%%%%%%
\begin{equation}
W_{ii}=-W_{00} \hbox{~~~for $i=1,2,3$}.
\end{equation}
%%%%%%%%%%%%%%%%%%%%%%%%%%%%%%%%%%%%%%%%%%%%%%%%%%%%%%%%%%%%%%%%%%%%%%%%%%%%%%%
Since the normalization of $W$ was absorbed in $x$, we adopt $W_{00}\!=\!1$
and the diagonal elements of $W$ are determined to be $\{1,-1,-1,-1\}$.
The remaining extremal conditions are
Eqs. (\ref{eq: equations by Bob's local filter}) and 
(\ref{eq: equations by Alice's local filter}), those are
%%%%%%%%%%%%%%%%%%%%%%%%%%%%%%%%%%%%%%%%%%%%%%%%%%%%%%%%%%%%%%%%%%%%%%%%%%%%%%%
\begin{equation}
\left\{
\begin{array}{l}
\displaystyle
\sum_{lm}\langle\phi_l|I\otimes\sigma_i|\tilde\phi_m\rangle W_{ml}\lambda_l=0
\cr
\displaystyle
\sum_{lm}\langle\phi_l|\sigma_i\otimes I|\tilde\phi_m\rangle W_{ml}\lambda_l
=0.
\end{array}
\right.
\label{eq: equations for W}
\end{equation}
%%%%%%%%%%%%%%%%%%%%%%%%%%%%%%%%%%%%%%%%%%%%%%%%%%%%%%%%%%%%%%%%%%%%%%%%%%%%%%%
On the other hand, by virtue of 
$F_A^\dagger \tilde F_A=F_A^\dagger\sigma_2 F_A^* \sigma_2=I$ 
\cite{Cen02a,Cen02b,Verstraete01a}
(we put $\det F_A\!=\!\det F_B\!=\!1$), we find
%%%%%%%%%%%%%%%%%%%%%%%%%%%%%%%%%%%%%%%%%%%%%%%%%%%%%%%%%%%%%%%%%%%%%%%%%%%%%%%
\begin{eqnarray*}
\langle\phi_l|I\otimes\sigma_i|\tilde\phi_l\rangle&=&
\langle e_l|F_A^\dagger \tilde F_A \otimes F_B^\dagger \sigma_i \tilde F_B
|\tilde e_l\rangle \cr
&=&\hbox{Tr}(I \otimes F_B^\dagger \sigma_i \tilde F_B)|e_l\rangle\langle e_l|
\cr
&=&\frac{1}{2}\hbox{Tr} \sigma_i \sigma_2 F_B^* \sigma_2 F_B^\dagger 
=\frac{1}{2}\hbox{Tr} \sigma_i=0,
\end{eqnarray*}
%%%%%%%%%%%%%%%%%%%%%%%%%%%%%%%%%%%%%%%%%%%%%%%%%%%%%%%%%%%%%%%%%%%%%%%%%%%%%%%
and therefore, 
Eqs.\ (\ref{eq: equations for W}) are the set of linear equations
only for the off-diagonal elements of $W$.
As the result, $W_{m\ne l}\!=\!0$
obviously satisfy Eqs.\ (\ref{eq: equations for W}).
After all,
%%%%%%%%%%%%%%%%%%%%%%%%%%%%%%%%%%%%%%%%%%%%%%%%%%%%%%%%%%%%%%%%%%%%%%%%%%%%%%%
\begin{equation}
W=\hbox{diag}\{1,-1,-1,-1\}
\label{eq: matrix W}
\end{equation}
%%%%%%%%%%%%%%%%%%%%%%%%%%%%%%%%%%%%%%%%%%%%%%%%%%%%%%%%%%%%%%%%%%%%%%%%%%%%%%%
is a solution of all extremal conditions.
Since $W^E\!=\!U\sqrt{\Pi}W\sqrt{\Pi}U^\dagger$,
we finally obtain
a relation between $\varrho$ and $\sigma$ in an analytical form as
%%%%%%%%%%%%%%%%%%%%%%%%%%%%%%%%%%%%%%%%%%%%%%%%%%%%%%%%%%%%%%%%%%%%%%%%%%%%%%%
\begin{eqnarray}
R^E&=&\Gamma+x\Delta^E \cr
\Delta^E&=&\big[(U\sqrt{\Pi}W\sqrt{\Pi}U^\dagger) \circ G\big]
\label{eq: relation in eigenstates}
\end{eqnarray}
%%%%%%%%%%%%%%%%%%%%%%%%%%%%%%%%%%%%%%%%%%%%%%%%%%%%%%%%%%%%%%%%%%%%%%%%%%%%%%%
with $G$ by Eq.\ (\ref{eq: matrix G}), $W$ by Eq.\ (\ref{eq: matrix W}),
and $\Pi$ by Eq.\ (\ref{eq: matrix Q}).
It should be noted that the phases in $U$, as well as $V$, should be
chosen so that $\langle \phi_i|\tilde \phi_j\rangle\!=\!\delta_{ij}$.
In order to avoid the complication,
it is convenient to express it in the Wootters basis, that is
%%%%%%%%%%%%%%%%%%%%%%%%%%%%%%%%%%%%%%%%%%%%%%%%%%%%%%%%%%%%%%%%%%%%%%%%%%%%%%%
\begin{eqnarray}
R^\Lambda&=&\Lambda+x\Delta^\Lambda \cr
\Delta^\Lambda&=&\sqrt{\Pi}U^\dagger
\big[(U\sqrt{\Pi}W\sqrt{\Pi}U^\dagger) \circ G\big]U\sqrt{\Pi}.
\label{eq: relation in Wootters}
\end{eqnarray}
%%%%%%%%%%%%%%%%%%%%%%%%%%%%%%%%%%%%%%%%%%%%%%%%%%%%%%%%%%%%%%%%%%%%%%%%%%%%%%%
In this expression, any $U$ satisfying
$U\sqrt{Q}\Lambda\sqrt{Q}U^\dagger\!=\!\Gamma$ can be used safely.
From the relation, it can be seen that the non-orthogonality of the Wootters
basis plays an important role.
\par
%%%%%%%%%%%%%%%%%%%%%%%%%%%%%%%%%%%%%%%%%%%%%%%%%%%%%%%%%%%%%%%%%%%%%%%%%%%%%%%
The trace of the matrix $\delta\!=\!\sum_{ij}\Delta^E_{ij}|i\rangle\langle j|$
is confirmed to be zero as
%%%%%%%%%%%%%%%%%%%%%%%%%%%%%%%%%%%%%%%%%%%%%%%%%%%%%%%%%%%%%%%%%%%%%%%%%%%%%%%
\begin{eqnarray}
\hbox{Tr}\Delta^E
&=&\sum_i [U\sqrt{\Pi}W\sqrt{\Pi}U^\dagger]_{ii} \gamma_i 
=\hbox{Tr}U\sqrt{\Pi}W\sqrt{\Pi}U^\dagger \Gamma \cr
&=&\hbox{Tr}W\Lambda
=\lambda_0-\lambda_1-\lambda_2-\lambda_3=0,
\end{eqnarray}
%%%%%%%%%%%%%%%%%%%%%%%%%%%%%%%%%%%%%%%%%%%%%%%%%%%%%%%%%%%%%%%%%%%%%%%%%%%%%%%
and $\delta$ obviously plays the role of the direction {\it normal}
to the boundary surface at the position of $\sigma$,
when $S(\varrho||\sigma)$ is regarded as the distant measure between states.
Which side does the entangled region spread?
For an infinitesimally small $|x|$,
by treating $\Delta^\Lambda$ as the perturbation,
the diagonal elements of $\Delta^\Lambda$ contribute
to the concurrence of $\varrho$ in the lowest order of $|x|$, and then
%%%%%%%%%%%%%%%%%%%%%%%%%%%%%%%%%%%%%%%%%%%%%%%%%%%%%%%%%%%%%%%%%%%%%%%%%%%%%%%
\begin{eqnarray*}
C(\varrho)&\sim&x\Delta C \cr
\Delta C&=&\hbox{Tr}\Delta^\Lambda W
=\sum_{ij}\big| \big[U\sqrt{\Pi}W\sqrt{\Pi}U^\dagger\big]_{ij}\big|^2 G_{ij}
>0,
\label{eq: change of concurrence}
\end{eqnarray*}
%%%%%%%%%%%%%%%%%%%%%%%%%%%%%%%%%%%%%%%%%%%%%%%%%%%%%%%%%%%%%%%%%%%%%%%%%%%%%%%
since $G_{ij}\!>\!0$ and
$U\sqrt{\Pi}W\sqrt{\Pi}U^\dagger$ is not zero operator (it is invertible).
Therefore, the entangled region always spreads in $x\!>\!0$.
When $\sigma$ is full rank, we can always find the positive $\varrho$
for some infinitesimally small $|x|$.
This implies that there are both entangled states 
and disentangled states in the vicinity of $\sigma$.
Therefore, the full rank $\sigma$ with $C\!=\!0$ is the true boundary state
separating the entangled and disentangled region.
Further, $E_R(\varrho)$ for $|x|\!\ll\!1$ is also calculated as
%%%%%%%%%%%%%%%%%%%%%%%%%%%%%%%%%%%%%%%%%%%%%%%%%%%%%%%%%%%%%%%%%%%%%%%%%%%%%%%
\begin{equation}
S(\varrho||\sigma)\sim x\hbox{Tr}(Z+I)\delta=
x^2\hbox{Tr}W\Delta^\Lambda=x^2 \Delta C.
\end{equation}
%%%%%%%%%%%%%%%%%%%%%%%%%%%%%%%%%%%%%%%%%%%%%%%%%%%%%%%%%%%%%%%%%%%%%%%%%%%%%%%
\par
%%%%%%%%%%%%%%%%%%%%%%%%%%%%%%%%%%%%%%%%%%%%%%%%%%%%%%%%%%%%%%%%%%%%%%%%%%%%%%%
It is important to note that there are entangled states satisfying 
Eqs. (\ref{eq: equations by Bob's local filter}),
(\ref{eq: equations by Alice's local filter}), and
(\ref{eq: equations by subspace}) for which $\sigma$ is not optimal.
However, the solution we obtained are not the case:
$\sigma$ is indeed optimal for $\varrho\!=\!\sigma+x\delta$ with any
$x\!\ge\!0$.
This can be proven as follows.
The equations we solved take into account any
small deviation from $\sigma$ along the boundary surface, and the solutions
are ensured to be extremal along any path on the boundary.
This extremeness does not immediately imply the
global minimum of $S(\varrho||\sigma)$ among the all disentangled states
by itself.
However, for our solution, $S(\varrho||\sigma+x\delta)\!>\!S(\varrho||\sigma)$
holds for $x\!<\!0$ (in the vicinity of $\sigma$),
and hence the relative entropy inside the boundary is larger than that on the
boundary.
In this situation, considering the semi-continuity and convexity of the
relative entropy \cite{Ohya93a},
it can be shown that the extremeness along the boundary must be local minimum
among the disentangled states around $\sigma$, and it must be global minimum \cite{Vedral98a}.
Further, the assumption we first made that $\varrho$ has $\sigma$ of
full rank is also justified in the same reason.
\par
%%%%%%%%%%%%%%%%%%%%%%%%%%%%%%%%%%%%%%%%%%%%%%%%%%%%%%%%%%%%%%%%%%%%%%%%%%%%%%%
It should be noted further about the uniqueness of $W$.
(if it is not unique,
there might exist entangled states other than those we obtained,
for which $\sigma$ is optimal).
Since $W$ is Hermitian, the number of real parameters of the off-diagonal
elements is 12, and the number of
linear Eqs.\ (\ref{eq: equations for W}) is also 12.
Therefore, Eq.\ (\ref{eq: matrix W}) is a unique solution unless
some of Eqs.\ (\ref{eq: equations for W}) are dependent to each other.
We have numerically generated random $\sigma$'s and confirmed that
Eqs.\ (\ref{eq: equations for W}) are always independent.
However, such Monte Carlo type calculations cannot pick up the
states of measure zero, such as those having some special symmetry,
and the other solutions $\varrho$ have not been completely excluded.
Further, it is an open question whether the solution we obtained
is enough to express whole entangled states
(does $\sigma+|x|\delta$ span whole entangled states, when $\sigma$ moves
on the boundary?)
\par
%%%%%%%%%%%%%%%%%%%%%%%%%%%%%%%%%%%%%%%%%%%%%%%%%%%%%%%%%%%%%%%%%%%%%%%%%%%%%%%
In so far, we solely discussed the case that $\sigma$ is full rank. 
In the case of $\sigma$ with a lower rank (denoted by $\sigma_L$),
let us consider a sequence constituted by all full rank $\sigma_n$'s
converging as $\sigma_n\!\rightarrow\!\sigma_L$.
We can always find such sequences because there always exist
full rank boundary states in the vicinity of $\sigma_L$
(if $\lambda_k$ of $\sigma_L$ is zero, adding infinitesimally small portion
of $|\phi_k\rangle\langle\phi_k|$ to $\sigma_L$).
In this sequence, $\varrho_n$ obtained from
Eq.\ (\ref{eq: relation in eigenstates}) also converge as
$\varrho_n\!\rightarrow\!\varrho_\infty$ (since $G_{ij}$ is non-singular).
Then, the continuity of the relative entropy of entanglement shown
in \cite{Donald99a} ensures that 
$E_R(\varrho_\infty)\!=\!S(\varrho_\infty||\sigma_L)$,
and $\sigma_L$ is one of the closest disentangled states
of $\varrho_\infty$.
After all, Eq.\ (\ref{eq: relation in eigenstates}) must hold
under such limiting sequences even for low rank $\sigma_L$.
\par
%%%%%%%%%%%%%%%%%%%%%%%%%%%%%%%%%%%%%%%%%%%%%%%%%%%%%%%%%%%%%%%%%%%%%%%%%%%%%%%
Finally, we show some simple examples
of Eq.\ (\ref{eq: relation in Wootters}).
The Wootters basis can be easily constructed 
by $|\phi_i\rangle\!=\!(F_A\!\otimes\!F_B)|e_i\rangle$.
For $F_A\!=\!F_B\!=\!I$, $\sigma$ becomes a Bell diagonal state,
and $|\phi_i\rangle\!=\!|e_i\rangle$ constitute the orthogonal set.
Then, $Q\!=\!\Pi\!=\!U\!=\!I$ and $\Gamma\!=\!\Lambda$, and we obtain
%%%%%%%%%%%%%%%%%%%%%%%%%%%%%%%%%%%%%%%%%%%%%%%%%%%%%%%%%%%%%%%%%%%%%%%%%%%%%%%
\begin{eqnarray*}
\Delta^\Lambda&=&\hbox{diag}
\big\{1/2,-\lambda_1,-\lambda_2,-\lambda_3\big\} \cr
R^\Lambda&=&\hbox{diag}
\big\{\frac{1+x}{2},(1-x)\lambda_1,(1-x)\lambda_2,(1-x)\lambda_3\big\}.
\end{eqnarray*}
%%%%%%%%%%%%%%%%%%%%%%%%%%%%%%%%%%%%%%%%%%%%%%%%%%%%%%%%%%%%%%%%%%%%%%%%%%%%%%%
For the choice of $F_A\!=\!F_B\!=\!\hbox{diag}\{t,1/t\}$,
$|e_0\rangle\!=\!i|\phi^+\rangle$, and
$|e_l\rangle\!=\! (I \otimes \sigma_l)|\phi^+\rangle$,
where $|\phi^+\rangle\!=\!(00\rangle+|11\rangle)/\sqrt{2}$, we have
%%%%%%%%%%%%%%%%%%%%%%%%%%%%%%%%%%%%%%%%%%%%%%%%%%%%%%%%%%%%%%%%%%%%%%%%%%%%%%%
\begin{equation}
Q=\Pi^*=\frac{1}{2}\left(
\begin{array}{cccc}
t^4+1/t^4 &&& -it^4+i/t^4\cr
& 1 && \cr
&& 1 & \cr
it^4-i/t^4 &&& t^4+1/t^4 
\end{array}
\right),
\end{equation}
%%%%%%%%%%%%%%%%%%%%%%%%%%%%%%%%%%%%%%%%%%%%%%%%%%%%%%%%%%%%%%%%%%%%%%%%%%%%%%%
and the limiting sequence of
$\lambda_1\!=\!\lambda_2\!\rightarrow\! 0$ and
$\lambda_3\!\rightarrow\!\lambda_0$
gives $\Gamma=\lambda_0\hbox{diag}\{ t^4,0,0,1/t^4\}$, and
%%%%%%%%%%%%%%%%%%%%%%%%%%%%%%%%%%%%%%%%%%%%%%%%%%%%%%%%%%%%%%%%%%%%%%%%%%%%%%%
\begin{eqnarray}
\Delta^\Lambda&\propto&\hbox{diag}\big\{1,0,0,-1\big\} \cr
R^\Lambda&=&\hbox{diag}\big\{1+x',0,0,1-x'\big\}.
\end{eqnarray}
%%%%%%%%%%%%%%%%%%%%%%%%%%%%%%%%%%%%%%%%%%%%%%%%%%%%%%%%%%%%%%%%%%%%%%%%%%%%%%%
These examples reproduce the correct relation between Bell diagonal
$\varrho$ and $\sigma$ \cite{Vedral97a}, and between
pure state and its closest disentangled state of rank 2 \cite{Vedral98a},
respectively.
\par
%%%%%%%%%%%%%%%%%%%%%%%%%%%%%%%%%%%%%%%%%%%%%%%%%%%%%%%%%%%%%%%%%%%%%%%%%%%%%%%
In summary,
we provided an analytical formula connecting $\sigma$ and $\varrho$
in two qubits
[Eq.\ (\ref{eq: relation in eigenstates})].
Using this formula, when $\sigma$ with $C\!=\!0$ is given, $\varrho$
for which $S(\varrho||\sigma)$ is minimum can be obtained analytically.
Further, this formula naturally defines the vector ($\delta$)
normal to the entangle-disentangle boundary surface.
The normal vector is uniquely determined
by $\sigma$ in almost all cases.
\par
%%%%%%%%%%%%%%%%%%%%%%%%%%%%%%%%%%%%%%%%%%%%%%%%%%%%%%%%%%%%%%%%%%%%%%%%%%%%%%%
This work was supported by CREST, Japan Science and Technology Corporation
(JST).
%%%%%%%%%%%%%%%%%%%%%%%%%%%%%%%%%%%%%%%%%%%%%%%%%%%%%%%%%%%%%%%%%%%%%%%%%%%%%%%
%%
%%%%%%%%%%%%%%%%%%%%%%%%%%%%%%%%%%%%%%%%%%%%%%%%%%%%%%%%%%%%%%%%%%%%%%%%%%%%%%%

%%%%%%%%%%%%%%%%%%%%%%%%%%%%%%%%%%%%%%%%%%%%%%%%%%%%%%%%%%%%%%%%%%%%%%%%%%%%%%%
%%
%%%%%%%%%%%%%%%%%%%%%%%%%%%%%%%%%%%%%%%%%%%%%%%%%%%%%%%%%%%%%%%%%%%%%%%%%%%%%%%
\end{document}